\documentclass[12pt]{article}
\usepackage[english]{babel}
\usepackage[letterpaper,top=2cm,bottom=2cm,left=3cm,right=3cm,marginparwidth=1.75cm]{geometry}
\usepackage{amsmath,amssymb,amsthm,bm,bbm,amsfonts,mathtools,thmtools} 
\usepackage{dsfont} 
\usepackage{graphicx}
\usepackage[authoryear, round]{natbib}
\usepackage[colorlinks=true, allcolors=blue]{hyperref}
\usepackage[nice]{nicefrac} 
\usepackage{xifthen}
\usepackage{xspace}
\usepackage{xparse}
\usepackage{xstring}
\usepackage{xspace}
\usepackage{float}
\usepackage{array}
\usepackage{inconsolata}

 \usepackage {cleveref}

\newcommand{\eqd}{\overset{d}{=}}
\newcommand{\given}{\mid}
\newcommand{\eps}{\epsilon}

\newcommand{\seqY}{$\{Y_1, \dots, Y_t \}$ }

\newcommand{\Fbar}{\overline{F}}

\newcommand{\VAR}{\textup{VaR}\xspace}
\newcommand{\EVAR}{\textup{EVaR}\xspace}
\newcommand{\ES}{\textup{ES}\xspace}
\newcommand{\MES}{\textup{MES}\xspace}
\newcommand{\QMES}{\textup{QMES}\xspace}
\newcommand{\XMES}{\textup{XMES}\xspace}

\newcommand{\xitilde}{\widetilde{\xi}\xspace}
\newcommand{\xihat}{\widehat{\xi}\xspace}
\newcommand{\xibar}{\overline{\xi}\xspace}
\newcommand{\xitildestar}{\widetilde{\xi}^{\star}\xspace}
\newcommand{\xihatstar}{\widehat{\xi}^{\star}\xspace}
\newcommand{\xibarstar}{\overline{\xi}^{\star}\xspace}
\newcommand{\qhatstar}{\widehat{q}^{\star}\xspace}
\newcommand{\qhat}{\widehat{q}\xspace}
\newcommand{\XMEStilde}{\widetilde{\textup{XMES}}\xspace}
\newcommand{\XMEShat}{\widehat{\textup{XMES}}\xspace}
\newcommand{\QMEShat}{\widehat{\textup{QMES}}\xspace}
\newcommand{\XMEStildestar}{\widetilde{\textup{XMES}^{\star}}{\xspace}}
\newcommand{\XMEShatstar}{\widehat{\textup{XMES}^{\star}}{\xspace}}
\newcommand{\QMEShatstar}{\widehat{\textup{QMES}^{\star}}{\xspace}}

\newcommand{\MESbarstar}{\overline{\textup{MES}}^{\star}\xspace}

\makeatletter
\newcommand{\@lcrx}[4][{-1}]{
          \IfEq{#1}{-1}{\left #2 {{{{#3}}}} \right #4}{
   	\IfEq{#1}{0}{#2 {{{{#3}}}} #4}{
	\IfEq{#1}{1}{\bigl #2 {{{{#3}}}} \bigr #4}{
	\IfEq{#1}{2}{\Bigl #2 {{{{#3}}}} \Bigr #4}{
	\IfEq{#1}{3}{\biggl #2 {{{{#3}}}} \biggr #4}{
	\IfEq{#1}{4}{\Biggl #2 {{{{#3}}}} \Biggr #4}{
    \GenericWarning{"4th argument to @lcrx must be -1, 0, 1, 2, 3, or 4"}
    }}}}}}}  
\newcommand{\lcrx}[4][{-2}]{
	\IfEq{#1}{-2}{\mathchoice{\@lcrx[-1]{#2}{#3}{#4}}
				 {\@lcrx[0]{#2}{#3}{#4}}
				 {\@lcrx[0]{#2}{#3}{#4}}
				 {\@lcrx[0]{#2}{#3}{#4}}}
		     {\@lcrx[#1]{#2}{#3}{#4}}}
\makeatother

\newcommand{\iid}{\textrm{i.i.d.}\@\xspace}
\newcommand{\D}[2]{\ensuremath{\frac{\partial #1}{\partial #2}}}
\newcommand{\convd}{\overset{d}{\to}}
\newcommand{\ind}{\mathds{1}} 
\def\argmin{\operatornamewithlimits{arg\,min}}
\newcommand{\EE}{\mathbb{E}}	
\newcommand{\reals}{\ensuremath{\mathbb{R}}}
\newcommand{\st}{\,:\,}
\newcommand{\distNorm}{\mathcal{N}}

\newcommand{\floor}[2][{-2}]{\lcrx[#1] \lfloor {#2} \rfloor}
\newcommand{\virgolette}[1]{``#1''}

\title{Tail Risk and Systemic Risk Estimation of Cryptocurrencies: an
Expectiles and Marginal Expected Shortfall based approach}
\author{Andrea Teruzzi\footnote{andrea.teruzzi@studbocconi.it}}
\date{}
\begin{document}
\maketitle
\begin{abstract}
This elaborate investigates the issue related to the quantification of the tail risk of cryptocurrencies. The statistical methods used in the study are those concerning recent developments of Extreme Value Theory (EVT) for weakly dependent data. This research proposes an expectile-based approach for the assessment of the tail risk of dependent data. Expectile is a summary statistic that generalizes the concept of mean, as the quantile generalizes the concept of median.  We present the empirical findings for a dataset of cryptocurrencies. We propose a method for dynamically evaluating the level of the expectiles by estimating the level of the expectiles of the residuals of a heteroscedastic regression, such as a GARCH model. Finally, we introduce the Marginal Expected Shortfall (MES) as a tool for measuring the marginal impact of single assets on systemic shortfalls. In our case of interest, we are focused on the impact of a single cryptocurrency on the systemic risk of the whole cryptocurrency market. In particular, we present an expectile-based MES for dependent data.
\end{abstract}
\section{Introduction}
In the financial industry, risk is an essential concept. It can be defined as the potential loss or gain associated with an investment decision. Risk measures such as the Value at Risk (\VAR) and the Expected Shortfall (\ES) are often used are used to quantify the riskiness of an asset. From a statistical point of view, the \VAR of level $\tau$ is the quantile $q_{\tau}$ of the same level, while the \ES of level $\tau$ is the average between all the observations after $q_{\tau}$. Firstly introduced in \cite{newey_1987}, expectiles asymmetrically generalize the concept of the mean in the same way the quantiles generalize the concept of the median. Given the random variable $X$, we define the expectiles $\xi(X)$ of level $\tau$ as the minimizer of the following  asymmetric and quadratic loss function
\begin{equation}
    \xi_\tau(X) = \argmin_{x \in \reals} \tau\EE [(X-x)^2_+] + (1-\tau)\EE [(X-x)^2_-]  
    \label{eq:exp}
\end{equation}
for $\tau \in (0,1)$. \EVAR is known to be the only coherent and elicitable risk measure \citep{bellini_2017} since the \VAR fails to be coherent, and the \ES fails to be elicitable \citep{gneiting_2011}. According to \cite{artzner_99}, a risk measure is said to be coherent when it satisfies the axioms of translation invariance, subadditivity, positive homogeneity, and monotonicity. Those properties are particularly useful as they grant good economic properties of the estimates. For instance, subadditivity is very important for asset managers, as it is closely related to the concept of diversification. The essential idea behind this property is that \virgolette{\textit{a merger does not create extra risk}}. Speaking practically, we would like to have a risk measure that does not give some extra risk for an extra asset in the portfolio, except for the amount of risk associated with the single new asset. Then, We say a risk measure to be elicitable if it can be defined as the minimizer of the expected value of a certain scoring function.  This property is desirable for statistical reasons, as it naturally provides a backtesting score for the estimates. \par
However, critics of the \EVAR underline the absence of an economic interpretation for the expectiles, as there is for the \VAR and the \ES. \EVAR has a clear and precise economic meaning; however, we need to introduce the concept of acceptance set of a risk measure to exploit it. We define the acceptance set of a risk measure $\rho$ as follows
     \begin{equation}
     \mathcal{A}_{\rho} = \{L | \rho(L) \leq 0\}.
     \end{equation}
The acceptance set can be seen as the decision criterion for which it is worth taking a particular financial position. For the VaR, the acceptance set is 
\begin{equation}
    \mathcal{A}_{\VAR_\tau}  =\Big\{X\Big|\frac{P(X>0)}{P(X<0)} \geq \frac{1-\tau}{\tau}\Big\}, \label{eq:as_var}
\end{equation}    
while for the \EVAR is
\begin{equation}
            \mathcal{A}_{\EVAR_\tau} =\Big\{X\Big|\frac{\EE(X_+)}{\EE(X_-)} \geq \frac{1-\tau}{\tau}\Big\} \label{eq:as_evar}
\end{equation}
\Cref{eq:as_var} and \Cref{eq:as_evar} have both a clear and succinct financial meaning. For the $\VAR_\tau$ a financial position is acceptable if the ratio of the probability of a gain over the probability of a loss is greater than a certain threshold depending on $\tau$. Similarly, for the $\EVAR_{\tau}$, a position is acceptable if the ratio between the expected gain over the expected loss is greater than a certain threshold depending on $\tau$. Therefore, the \EVAR is a coherent, elicitable and financially meaningful risk measure.

\section{Statistical model}
Let \seqY to be a strictly stationary and beta-mixing time series representing a negative financial position, having continuous cumulative distribution function $F$. We interpret a large value of $Y$ as a large loss. Furthermore, we assume that $F$ is regularly varying at index $-1/\gamma$, namely $\Bar{F}(ty)/\Bar{F}(ty) \rightarrow -1/\gamma$ for $t\rightarrow\infty$. This assumption implies $\gamma>0$. We additionally assume $\EE(|Y_{-}|)$ with $\gamma<1$. The assumptions reflect a situation of heavy-tailed time series, but not too heavy in order to guarantee the existence of the first moment of $Y$. Consequently, the expectile $\xi_{\tau}$ is well defined for every $\tau$.  \par
The statistical framework follows the approach proposed by \cite*{padoan_2022}. The inference on extreme values of $\xi$ is based on a family of semiparametric estimators of the form $\hat{\xi}_{\tau_n'} = \hat{\xi}_{\tau_n'}(\xi_{\tau_n})$ with $\tau_n<\tau_n'$.  We introduce two different estimators for $\xi$ at the intermediate level $\tau$. The first one is called least asymmetrically weighted squares (LAWS) and can be seen as the empirical counterpart of \Cref{eq:exp}
\begin{equation}
\xitilde_{\tau_n} = \argmin_{\theta \in \reals} \sum_{t=1}^N \eta_{\tau_n}(Y_t-\theta),
\end{equation}
with $\eta_{\tau}(x) = |\tau-\ind\{x \leq 0\}|x^2$ the expectile check function.
The second intermediate estimator for $\xi_{\tau}$ is called quantile-based (QB), as it is founded on an asymptotic relation between the expectile and the quantile of the same level $\tau$. According to \cite{bellini_2014}
\[
\xihat_{\tau_n} = (\widehat{\gamma}^{-1}_n-1)^{-\widehat{\gamma}_n}\widehat{q}_{\tau_n}, \quad \textup{ for } \gamma<1,
\]
where $\widehat{q}_{\tau_n} = Y_{n-\floor{n(1-\tau_n)},n}$ is  the empirical quantile of level $\tau_n$. \par

Let $\tau_n'\rightarrow 1$ such that $n(1-\tau_n')\rightarrow c \in [0,\infty)$ as $n\rightarrow\infty$. Usually $\tau_n' = 1 - p_n$, where the exceedance probability $p_n$ is not greater than $1/n$ \citep{cai_2015}. We define an extreme estimator for $\xi_{\tau_n'}$ similar to the one proposed in \cite{weissman_1978}. Denoting the tail quantile function as $U:x\mapsto \inf\{y\in \reals \st 1/\Fbar(y)\geq x\}$, for $n$ large enough
\begin{equation}
    \frac{\xi_{\tau_n'}}{\xi_{\tau_n}} \approx \frac{q_{\tau_n'}}{q_{\tau_n}} = \frac{U((1-\tau_n')^{-1})}{U((1-\tau_n)^{-1})} \approx \bigg(\frac{1-\tau_n'}{1-\tau_n} \bigg)^{-\gamma}.
    \label{eq:proxy}
\end{equation}
From \Cref{eq:proxy}, we introduce a family of extrapolating estimators for the extreme expectile $\xi$ of level $\tau_n'$ given an intermediate level $\tau_n$
\begin{equation}
    \xibarstar_{\tau_n'} \equiv  \xibarstar_{\tau_n'}(\tau_n) = \bigg(\frac{1-\tau_n'}{1-\tau_n} \bigg)^{-\widehat{\gamma_n}}\xibar_{\tau_n},
\end{equation}
where $\widehat{\gamma_n}$ is an estimator for the tail index $\gamma$. Under the conditions of \cite*{padoan_2022}
\begin{equation}
\frac{\sqrt{n(1-\tau_n)}}{\log\{(1-\tau_n)/(1-\tau_n')\}}\bigg(\frac{\xibarstar_{\tau_n'}}{\xi_{\tau_n'}}-1\bigg) \convd \distNorm \bigg(\frac{\lambda_1}{1-\rho},  \gamma^{2}\bigg\{1+2\sum_{t=1}^{\infty}R_t(1,1) \bigg\}\bigg) \label{eq:bias_term}
\end{equation}
holds for
\begin{itemize}
    \item the LAWS intermediate estimator $\xibarstar_{\tau_n'} =\xitilde^{\star}_{\tau_n'}$
    \item the QB intermediate estimator $\xibarstar_{\tau_n'} =\xihat^{\star}_{\tau_n'}$.
\end{itemize}
To quantify the asymptotic variance and the bias term of  \Cref{eq:bias_term}, we apply the approach presented in \cite{drees_2003} for $\beta$-mixing time series. Let $w(\gamma, R)= \gamma^{2}\{1+2\sum_{t=1}^{\infty}R_t(1,1)\}$ be the asymptotic variance of our estimate. Consider big blocks of length $r_n$ and small blocks of length $l_n$. Then, define the following sequence of random variables:
\[
\widehat{Z}_j = \sum_{t=1+j l_n}^{r_n+j l_n} \ind \{\widehat{F}_n(Y_t) \geq \tau_n\} 
\]
for $j=0,\dots, m_n-1$, where $m_n=\floor{n/\ell_n}$, $\ell=r_n+l_n$ and $\widehat{F}_n$ the empirical distribution function of $Y$ using $n$ data. Finally, consider the estimator
\[
\widehat{w}_n(\gamma, R)= \frac{(\widehat{\gamma}_n^{H})^2}{r_n(1-\tau_n)} \Sigma_n
\]
with $\Sigma_n$ the sample variance of $\widehat{Z}$. Thus, we have the following confidence interval of level $(1-\alpha)$ for $\xi_{\tau_n'}$  
\[
\xibarstar \left(\frac{1-\tau_n}{1-\tau_n'}\right)^{
-\widehat{b}_n \pm z_{1-\alpha}\sqrt{\widehat{w}_n(\gamma, R)/\{n(1-\tau_n) \}}},
\]
where $\sqrt{k} \widehat{b}_n$ is an estimate of the bias term, which is $\lambda/(1-\rho)$.
We consider three different cases of confidence intervals derived from LAWS estimators:
\begin{itemize}
    \item LAWS-D-ADJ, when the bias term is taken into consideration,
\item LAWS-D, when the bias term is not taken into consideration,
\item LAWS-IID, when the asymptotic variance is calculated from the \iid case, leading to $w(\gamma, R)=\gamma^2$. 
\end{itemize}
Similar confidence intervals could be derived for the QB estimator leading to QB-D-ADJ, QB-D, and QB-IID. \par
The choice of the extreme level $\tau_n'$ should be able to have a particular financial meaning. That stands true in theory as well as in practice. However, from the point of view of a risk manager could be particularly insightful to select some fixed values. One viable option is selecting the largest reachable level of $\tau_n'$, which is $\tau_n' = 1 - 1/n$ according to \cite{cai_2015}. The alternative proposed by \cite*{padoan_2022} is to set $\tau_n'\equiv\tau_n'(\alpha_n)$ such that $\xi_{\tau_n'}\equiv\xi_{\tau_n'}(\alpha_n)\equiv q_{\alpha_n}$. When we deal with risk measures, this choice appears to be exceptionally reasonable because it is possible to take advantage of the larger interpretability of \VAR while retaining the coherency property of \EVAR. The level $\tau_n'(\alpha_n)$ is the one such that
\begin{equation}
\frac{1-\tau_n'(\alpha_n)}{1-\alpha_n} \rightarrow \frac{\gamma}{1-\gamma}
\end{equation}
as $n\rightarrow\infty$. From this relation, it follows the following estimator for large $n$
\begin{equation}
    \widehat{\tau}_n'(\alpha_n) = 1 - (1-\alpha_n)\frac{\widehat{\gamma}_n^{H}}{1-\widehat{\gamma}_n^{H}}. \label{eq:tau_hat}
\end{equation}

\section{Systemic risk}
In their broadest meaning, financial assets are not stand-alone entities but rather parts of a financial system. For example, stocks are usually not held singularly but rather in portfolios. Systemic risk is a term used in the financial and insurance industries, referring to the risk of a systemic event, such as a financial crisis, affecting multiple lines of business. \par
We define the Marginal Expected Shortfall (\MES) as the sensitivity of overall risk to its component $i$. Let $L^{FS} = \sum_{i=1}^{K}w_i L_i$ with $\sum_i^Kw_i=1$ and $L_i$ be the losses attributed to the overall risk and its component $i$ with weight $w_i$; we define the \MES of level $\alpha$ according to \cite{acharaya_2017}
\begin{equation}
    \MES_{\alpha}^i = \D{\ES_{\alpha}^{FS}}{w_i} = \EE ( L_i | L^{FS} > \VAR_{\alpha}^{FS} ),
\end{equation}
where $\VAR_{\alpha}^{FS}$ and $\ES_{\alpha}^{FS}$ are the overall value-at-risk and expected shortfall of level $\alpha$.The \MES measures how group $i$'s risk-taking adds to the overall risk. \MES at extreme quantile levels has been studied by \cite{cai_2015} in the form
\begin{equation}
\QMES_{X, \tau} = \EE( X | Y > q_{Y,\tau}) \quad \tau \in (0,1)
\end{equation}
where $X$ is the loss of the individual asset, $Y$ the aggregated loss and $q_{Y,\tau}$ the quantile of level $\tau$ of $Y$. \cite{daouia_2017} have proposed an expectile-based version of \MES for  $\tau$ close to 1,
\begin{equation}
\XMES_{X, \tau} = \EE( X | Y > \xi_{Y,\tau}) \quad \tau \in (0,1).
\end{equation}
Let us consider the extrapolating estimator $\MESbarstar_{\tau'} \equiv \MESbarstar_{\tau'}(\tau)$ for weakly-dependent time series similar to the one proposed for the \EVAR
\begin{equation}
    \MESbarstar_{X,\tau'_n} = \left(\frac{1-\tau'_n}{1-\tau_n}\right)^{-\overline{\gamma}_{X,n}} \frac{\sum_{t=1}^{n}X_t \ind\{X_t>0,Y_t> \overline{z}_{Y,\tau_n}\}}{\sum_{t=1}^{n}\ind\{Y_t> \overline{z}_{Y,\tau_n}\}}, 
\end{equation}
under the assumptions of \cite*{padoan_2022}
\begin{equation}
    \frac{\sqrt{n(1-\tau_n)}}{\log\{(1-\tau_n)/(1-\tau_n')\}}\bigg(\frac{\MESbarstar_{X,\tau_n'}}{\MES_{X,\tau_n'}}-1\bigg) \convd \distNorm \left( 0, \gamma_X^2\left\{1+2\sum_{t=1}^{\infty}R_{X,t}(1,1) \right\} \right)
\end{equation}
Substituting $\overline{z}_{Y,\tau_n}$ we obtain 
\begin{itemize}
    \item $\QMEShat_{X,\tau_n'}^{\star}$ when $\overline{z}_{Y,\tau_n} = \overline{q}_{Y,\tau_n}$,
    \item $\XMEStilde_{X,\tau_n'}^{\star}$ when $\overline{z}_{Y,\tau_n} = \xitilde_{Y,\tau_n}$,
    \item $\XMEShat_{X,\tau_n'}^{\star}$ when 
    $\overline{z}_{Y,\tau_n} = \xihat_{Y,\tau_n}$.
\end{itemize}
Asymptotic confidence intervals can be derived with the same procedure of the \EVAR.

\section{Application to cryptocurrency data}
\subsection{Dataset}
The proposed dataset regards Bitcoin, Ethereum, and Litecoin transactions from August 9, 2015, to July 6, 2021. The data were downloaded from Kaggle\footnote{https://www.kaggle.com/}.
\begin{figure}[!b]
    \centering
    \hspace*{-.47cm}  
    \includegraphics[width=1.05\textwidth,  keepaspectratio]{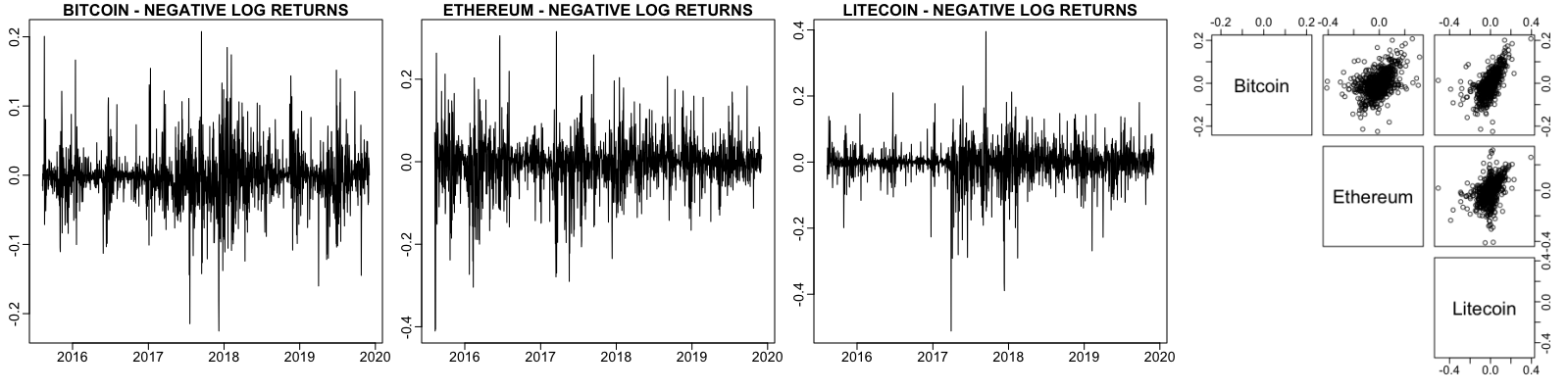}
    \caption{\textit{The first three plots show the negative log-returns of Bitcoin, Ethereum, and Litecoin. The fourth one shows three scatter plots representing the negative log-returns every coin against the other two.}}
    \label{fig:merged}
\end{figure}
These three coins represent three different archetypes of coins. Bitcoin and Ethereum are the only ones of their species, as they can be seen as the only large-cap and mid-cap cryptocurrencies. Differently, Litecoin is a small-cap coin, which is a prevalent feature within the crypto market. We consider the negative log-returns on the market prices of the coins to achieve a more regular series with mode approximately zero and without any evident drift. \Cref{fig:merged} shows the series and the three scatter plots of every coin against the other two. Only from a rough visual analysis it is possible to observe that the coins have a heavy right tail. Furthermore, a certain degree of dependence is evident by looking at the scatter plot in the upper right area among the pairs of cryptocurrencies.
\subsection{Estimation of Tail Expectiles} 
We also introduce estimators for extreme quantiles to compare them with the estimates of extreme expectiles. According to \cite{drees_2003}, the Weissman-type extreme quantile estimator of level $\tau'_n$ is
\begin{equation}
    \qhatstar_{\tau'_n} = \left( \frac{1-\tau'_n}{1- \tau_n}\right)^{-\gamma_n^H} \qhat_{\tau_n} 
\end{equation}
with $\tau_n<\tau'_n$. The most obvious choice for the intermediate level $\qhat_{\tau_n}$ is the empirical estimator for the quantile of level $\tau_n$, $Y_{n-\floor{n(1-\tau_n)},n}$. The asymptotic properties of this estimator have been shown by \cite{drees_2003}, in particular, we can use similar tools to the one introduced for extreme expectiles.  \par 
\begin{figure}[!ht]
    \centering
    \hspace*{-.66cm}  
    \includegraphics[width=1.08\textwidth,  keepaspectratio]{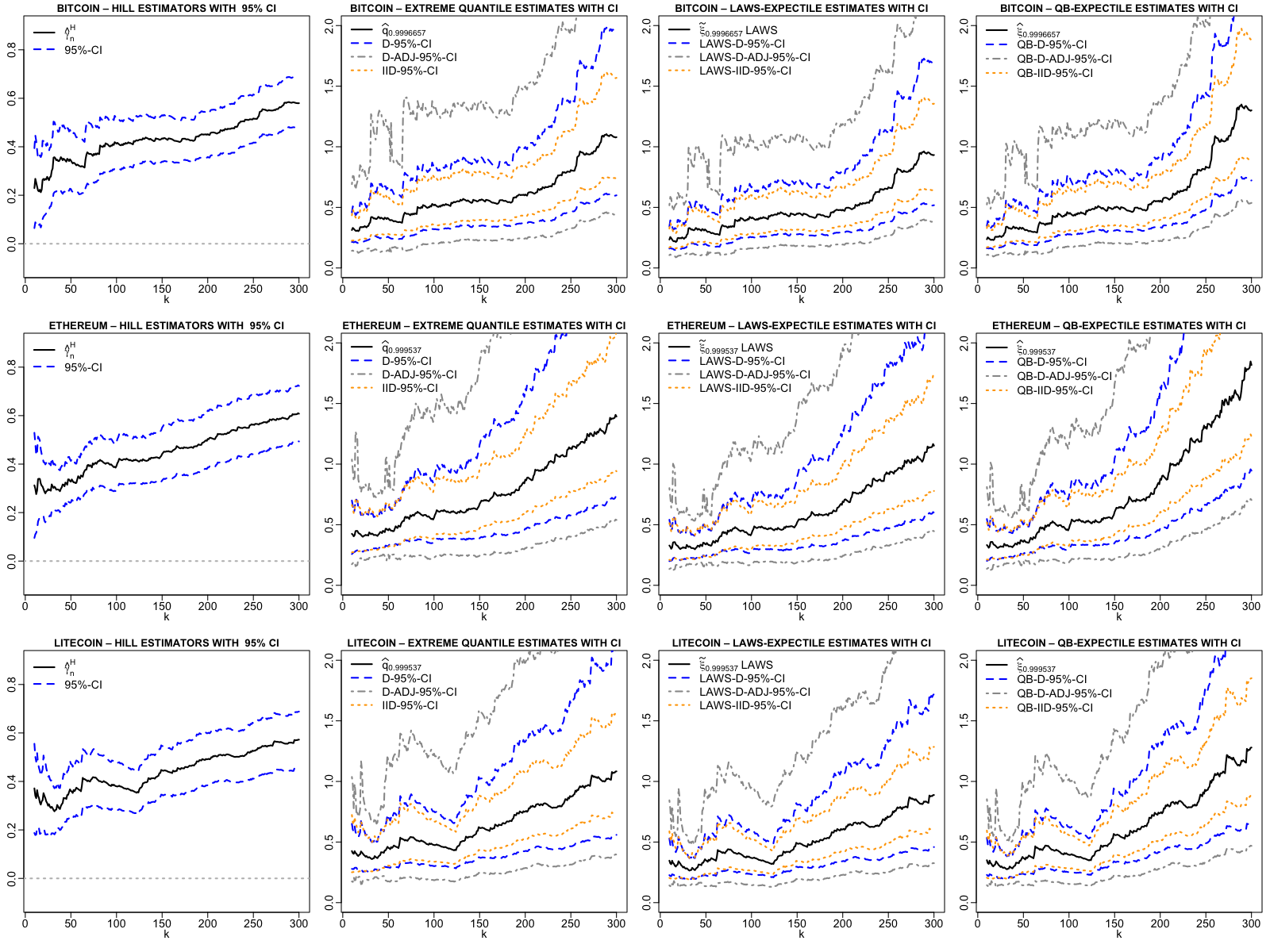}
    \caption{\textit{The first line of the lattice from left to right shows for Bitcoin: Hill estimator as a function of $k$; Weissman quantile estimator with D, D-ADJ, and IID 95\% confidence intervals as a function of $k$; LAWS-expectile estimator with D, D-ADJ, and IID 95\% confidence intervals as a function of $k$; QB-expectile estimator with D, D-ADJ, and IID 95\% confidence intervals as a function of $k$. The second and third lines show the same charts for Ethereum and Litecoin.}}
    \label{fig:static}
\end{figure}
The estimation of extreme expectiles and quantiles has been carried through the package \texttt{ExtremeRisks} \citep{ExtremeRisks}, which is present in \texttt{R} \citep{R}. The first thing we do is to set the value of $\tau'_n$ to be used for the analysis. We fix $p_n=1/n$ such that  $\tau'_n=1-p_n=0.999657$ for the Bitcoin time series and $\tau'_n=1-p_n=0.999537$ for Ethereum and Litecoin. Before evaluating the \EVAR and the \VAR of the selected cryptocurrencies, it is helpful to check the value of the tail index $\gamma$. The left column of the lattice in \Cref{fig:static} shows the Hill estimator of the tail index against $k$, which is the number of extreme observations to be taken into account for calculating the estimates. The estimates of Bitcoin and Ethereum are relatively stabilized between 100 and 150, while the estimates of $ \gamma $ for Litecoin are more wiggly, but the plot seems sufficiently stable around 125. Selecting $k=125$ for all three coins appears to be a wise choice. In this way, we obtain tail estimates approximately equal to 0.43, 0.41, and 0.40, respectively, for Bitcoin, Ethereum, and Litecoin. Those estimates are signs of a fat tail but with finite expectation, which is a crucial property for estimating $\xi$. Then, we need to choose the dimension of the big-block and small-block sequences for estimating the confidence intervals. A popular choice for geometric $\beta$-mixing time series is $l_n=\floor{ \log n}$  and $r_n=\floor{ \log^2 n}$. 
The last three columns of the lattice in \Cref{fig:static} shows the estimates for extreme quantile $\qhatstar_{\tau'_n}$, the LAWS estimator $\xitildestar_{\tau'_n}$ and the QB estimator $\xibarstar_{\tau'_n}$. As for the tail estimators, the plots are reasonably stable for Bitcoin and Ethereum, while the behavior of the estimates of Litecoin is more uncertain. The choice of $k=125$ is again reasonable, as all three cryptocurrencies seem stable enough in that area of the charts. Together with the point estimates, the plots show the 95\% confidence intervals depending on the value $k$. The independent intervals (IID) are the narrowest for all coins and point estimates. Indeed, these confidence intervals assume no clusters of extremal observations. 
\begin{table}[!hb]
\centering
\caption{\textit{Risk measure estimates with 95\%-D confidence intervals.}}
\renewcommand{\arraystretch}{1.3}
\begin{tabular}[!ht]{m{5em} m{9.3em} m{9.3em} m{9.3em}}
\hline
Estimate & Bitcoin & Ethereum & Litecoin\\
\hline
$\tau'_n$                & 0.999657             &  0.999537               & 0.999537              \\
$\widehat{\gamma}^H$     & 0.401 [0.289, 0.512] &  0.413 [0.310, 0.515]   & 0.397 [0.287, 0.507]  \\
$\qhatstar_{\tau'_n}$    & 0.495 [0.302, 0.811] &  0.597 [0.379, 0.940]   & 0.505 [0.310, 0.822]  \\
$\xitildestar_{\tau'_n}$ & 0.389 [0.238, 0.638] &  0.470 [0.299, 0.740]   & 0.402 [0.247, 0.654]   \\
$\xihatstar_{\tau'_n}$   & 0.421 [0.257, 0.690] &  0.516 [0.328, 0.812]   & 0.428 [0.263, 0.697]   \\
\hline
\label{tab:static}
\end{tabular}
\end{table}
In a certain sense, this assumption makes the sample more informative, making the confidence intervals less uncertain. In the same way, the temporal dependent confidence intervals (D) are more precise than the adjusted temporal dependent (ADJ-D). Then, the ADJ-D confidence intervals seem to be unreasonably broad for all estimates. The LAWS and the QB estimators have similar graphs; however, showing some consistently higher values for the QB estimates for the same value of $k$. Furthermore, the QB estimator shows a stronger drift for high values of $k$ (over 200) for all estimates, leading to more uncertain quantification of the risk measure. \par
From a financial point of view, Bitcoin, Ethereum, and Litecoin show similar and very risky behavior. \Cref{tab:static} shows the results of the analysis. It is important to remember that those \VAR and \EVAR estimates are calculated for daily return data. By looking at the results, the probability of losing half of the value of the investment from one day to another is small but tangible in a reasonably large time window. In particular, Ethereum is the cryptocurrency that presents the most significant marginal risk, followed by Bitcoin.
\subsection{Dynamic Estimation of Tail Expectiles} \label{sec:crypto_dyanamic}
We propose a dynamic extreme expectiles estimation using a method first proposed by \cite{mcneil_2000} for estimating extremal quantiles. The proposal is to use heteroscedastic regression models - such as ARCH and GARCH models - to estimate the innovation process's tail. The method proposed in \cite{mcneil_2000} is used for estimating extremal \VAR dynamically, while \cite*{padoan_2022} has introduced a similar method for dynamic extremal \EVAR. A general heteroscedastic regression can be written as
\begin{equation}
    Y_t = g(X_t)+\sigma(X_t)\epsilon_t
\end{equation}
where $g$ and $\sigma>0$ are unknown function to be estimated for obtaining predictions of $Y_t$ and $\epsilon_t$ is the innovation at time $t$. This equation includes classical autoregressive time-series models such as ARMA and GARCH. Assuming $\epsilon_t \eqd \epsilon$ for every $t$, using the positive homogeneity property of expectiles
\begin{equation}
    \xi_{\tau}(Y_t \given X_t) = g(X_t)+\sigma(X_t)\xi_{\tau}(\epsilon).
\end{equation}
This equation implies that given $g$ and $\sigma$, if it is possible to estimate $\xi_{\tau}(\eps)$, you can use this estimate for retrieving the value of $\xi_{\tau}(Y_t \given X_t)$. The general dynamic estimator will be in the form
\begin{equation}
    \xibar_{\tau}(Y_t \given X_t) = \overline{g}(X_t) + \overline{\sigma}(X_t)\xibar_{\tau}(\overline{\epsilon}),
\end{equation}
where $\overline{g}$ and $\overline{\sigma}$ are estimators of the heteroscedastic regression, and $\xibar_{\tau}(\overline{\epsilon})$ is the expectile estimator for the residuals of the regression $\overline{\epsilon}$. \cite*{padoan_2022} has shown that when $\xibar_{\tau}(\overline{\epsilon})$
is a LAWS or QB extrapolating estimator calculated using the Hill estimator for the tail index $\gamma$, the dynamic estimator $\xibar_{\tau}(Y_t \given X_t)$ is asymptotically Normal distributed with asymptotic bias and variance that can be quantified using asymptotic estimators. \par
\begin{table}[!ht]
\centering
\caption{\textit{Information criterion results for the first 1000 observations of Bitcoin.}}
\hspace*{-0.6cm}
\begin{tabular}[!ht]{m{2em} m{5.4em} m{5.4em} m{5.4em} m{5.4em} m{5.4em} m{5.4em}}
\hline 
&  \footnotesize{ ARMA(1,1) } & \footnotesize{ ARMA(1,0) }  & \footnotesize{ ARMA(1,1) }  & \footnotesize{ ARMA(1,2) } & \footnotesize{ARMA(1,1)}&  \footnotesize{\textbf{GARCH(1,1)}} \\
& \footnotesize{GARCH(1,1}) & \footnotesize{GARCH(1,1}) & \footnotesize{GARCH(2,1}) & \footnotesize{GARCH(2,1}) &         &             \\
\hline
AIC  & -\textbf{4.028} & -\textbf{4.028} & -4.026 & -4.026 & -3.927 & -4.024   \\
BIC  & -3.993 & -3.998 & -3.987 & -3.981 & -3.902 & -\textbf{3.999 }  \\
\hline
\label{tab:information}
\end{tabular}
\end{table}
Here we focus on applying this dynamic method to our cryptocurrency dataset. The first thing to do is to select the regression method. The main viable options concern heteroscedastic regression methods of the kind $\textup{ARMA}(p,q)+\textup{GARCH}(r,s)$. Simply looking at the plots of our data given by \ref{fig:merged}, it is clear that we shall put more emphasis on modeling the variance with a $\textup{GARCH}(r,s)$, rather than modeling the mean. To enhance the dynamic approach of the method, the models were trained using rolling windows of size $T=1000$. Namely, the estimated residuals at time $t+T$ are equivalent to $Y_{t+T}-\overline{Y}_{t+T}$, where $\overline{Y}_{t+T}$ is the prediction of the series using $\{Y_t, \dots, Y_{t+T-1} \}$ as regressors. To select the regression model, we evaluated the goodness of fit using standard information criteria, such as AIC and BIC. As an example, \Cref{tab:information} presents the results for the first 1000 observations of Bitcoin using several regression models. Ultimately, GARCH(1,1) was chosen because of the model's simplicity and the satisfactory goodness of fit. In addition, we put more emphasis on the distribution of the residuals to be heavy-tailed and \iid than on the predictive performances of the model. The regression model estimation has been carried through the package \texttt{rugarch} \citep{rugarch}. \par
\begin{figure}[!hb]
    \centering   
    \hspace*{-.66cm} 
    \includegraphics[width=1.08\textwidth,  keepaspectratio]{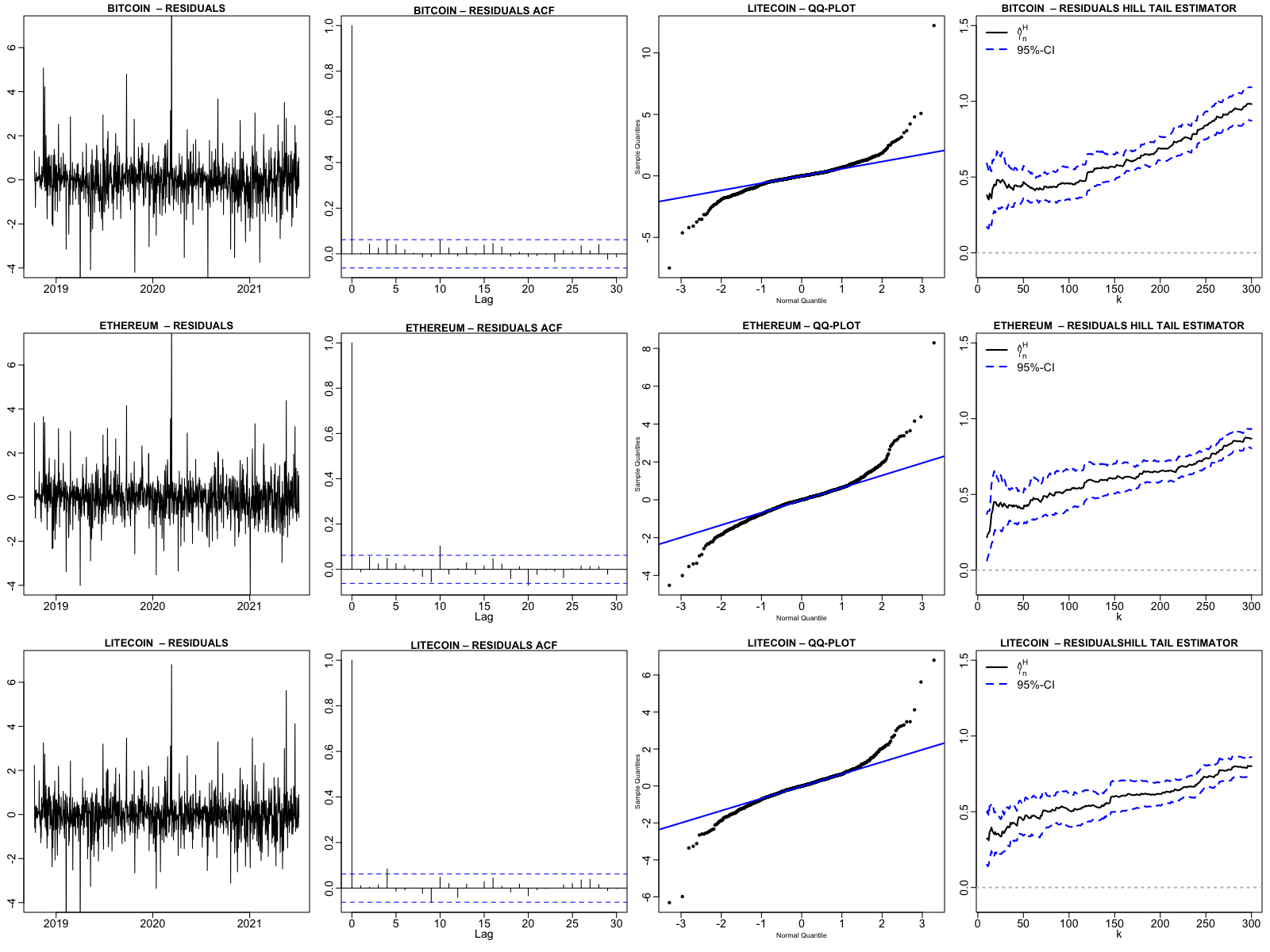}
    \caption{\textit{The first line of the lattice from left to right shows: the normalised residuals of the first 1000 observations of Bitcoin; the autocorrelation function of the residuals; a Q-Q plot against the Normal distribution of the residuals; the Hill estimator of the residuals as a function of $k$. The second and third lines show the same charts for Ethereum and Litecoin residuals.}}
    \label{fig:residuals}
\end{figure}
Before estimating the risk measure, it is a good practice to check the distribution of our model's residuals. The assumptions underlying our model require that the residuals are \iid distributed with a heavy right tail. The second column of the lattice in \Cref{fig:residuals} shows the autocorrelation function of the residuals of the cryptocurrencies. The estimates are almost always under the confidence interval represented by the dashed blue line; thus, we state quite comfortably that we do not have temporal dependence in our sample. At any rate, the amount of temporal dependence is negligible for all three coins. From the third and last columns, it is possible also to notice the presence of a heavy tail with a tail index of approximately equal to 0.5 for all the cryptocurrencies. Furthermore, a tail index smaller than 1 guarantees the existence of the first moment of the distribution and so expectiles to be well defined. The behavior of the residuals is in line with the assumptions of our model. \par
\begin{figure}[!hb]
    \centering
    \hspace*{-.66cm}
    \includegraphics[width=1.08\textwidth,  keepaspectratio]{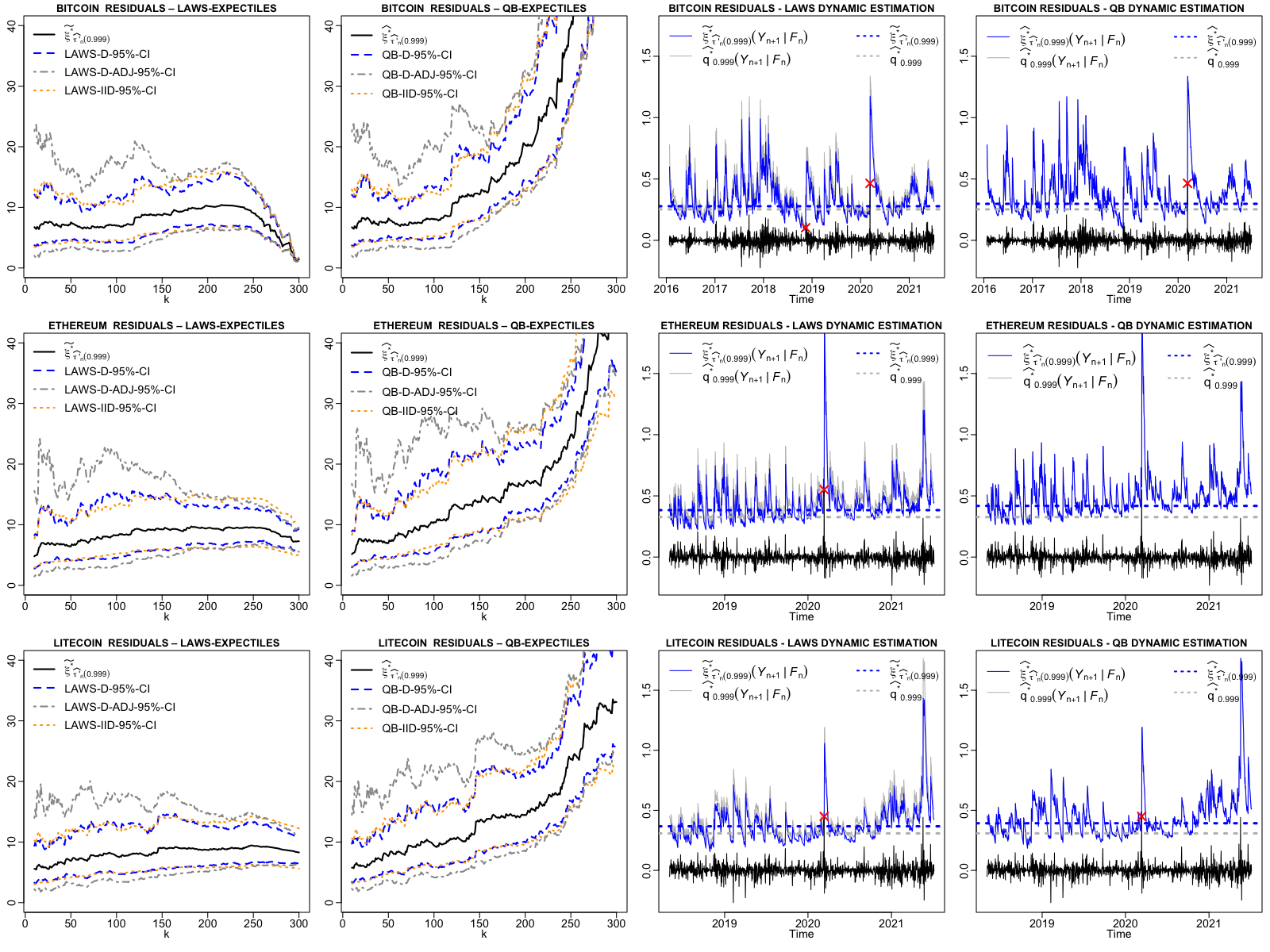}
    \caption{\textit{The first line of the lattice from left to right shows for Bitcoin: static LAWS expectiles for first 1000 observations with 95\% confidence intervals; static QB expectiles for first 1000 observations with 95\% confidence intervals; static and dynamic LAWS expectiles of level $\widehat{\tau}'_n(0.999)$ with static and dynamic quantile estimates of level $\tau'_n=0.999$; static and dynamic QB expectiles of level $\widehat{\tau}'_n(0.999)$ with static and dynamic quantile estimates of level $\tau'_n=0.999$. Red crosses represent exceedances over the dynamic level. The second and third lines show the same charts for Ethereum and Litecoin.}}
    \label{fig:dynamic}
\end{figure}

For this section, it seems reasonable to use the estimator $\widehat{\tau'}_n$ to have expectiles estimates that can be compared to quantiles estimates. Consequently, we achieve \EVAR estimates that can be interpreted and compared to \VAR estimates. \Cref{fig:dynamic} shows the results of the analysis. The first two columns of the lattice present \EVAR estimates using the LAWS and QB methods with 95\% confidence intervals for the first 1000 observations. Interestingly, the confidence regions defined by the D and the IID methods are almost overlying. That is another proof of the fact that the temporal dependence between residuals is extremely low. On the other side, ADJ-D  confidence intervals appear to be still somewhat too large. Furthermore, the LAWS estimates are much more stable than the QB estimates. Indeed, in the QB estimator plots is possible to see evident drifts starting from $k=200$, resulting in a more uncertain estimation. Selecting $k=125$ for all three coins again appears to be a wise choice. \par
The last two columns of the lattice show the result of the dynamic method. In particular, the first column exhibits the dynamic and static LAWS estimates of level $\widehat{\tau}'_n(0.999)$  and the static and dynamic extreme quantile estimates of level $\tau'_n=1-1/T=0.999$. On the other side, the second column shows QB estimates and extreme quantile estimates. As a first thing, we can see that the expectiles estimates and the quantile estimates are mostly overlying; to be specific, the QB expectiles and quantiles are almost completely overlapped. To evaluate the accuracy of the model, the plots present red crosses when the negative log returns exceed the level set by the dynamic prediction. For Bitcoin, we expect $(2990-T) \times (1-0.999) \approx 2.0$ exceedances above the $\VAR_{0.999}$, represented by the dynamic expectile $\xi_{\tau'_n(0.999)}$. For the LAWS estimates, we have two exceedances, while just one for the QB ones. For Ethereum and Litecoin we expect $(1578-T) \times (1-0.999) \approx 0.6$ exceedances. In practice, Ethereum has one exceedance for the LAWS estimates and zero for the QB estimates. For the Litecoin data, we have one exceedance over the dynamic LAWS expectiles and one for the QB ones. In summary, all six dynamic charts have a number of exceedances extremely close to the target number or even equal to the target. Specifically, the LAWS estimator seems to give better results than the QB one; however, it is difficult to generally state the overperformance of one method over the other one. Finally, even if the backtesting performances are excellent, the sample size of the dataset is too small to claim general results about the performance of the dynamic approach. Nevertheless, the proposed approach appears to be extremely interesting and definitely applicable to this cryptocurrency dataset.
\subsection{Estimation of MES} \label{sec:crypto_mes} 
For the calculation of the \XMES is necessary to find a market index of the cryptocurrency market coherent with the definition of overall risk given by \cite{acharaya_2017}.  There exist multiple options. The first and most natural way would have been to select an index representing the cryptocurrency market, as the S\&P500 and Dow Jones Industrial are for the stock market. However, this choice appeared somewhat unsatisfactory because the financial industry has no crypto market index considered as a reference for everybody. There exist several indexes provided by rating agencies and other financial institutions; nevertheless, they are not entirely recognized by all investors. Furthermore, the calculation of the indexes is sometimes sophisticated and complex, and so not wholly in line with the definition of \cite{acharaya_2017}.
\begin{figure}[!hb]
    \centering
    \hspace*{-.66cm}
    \includegraphics[width=1.08\textwidth,  keepaspectratio]{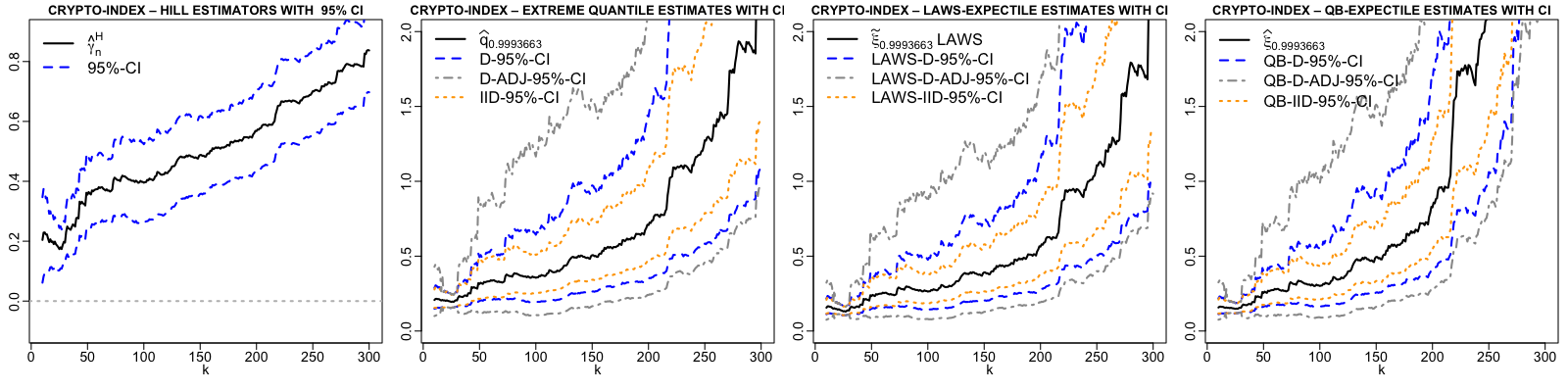}
    \caption{\textit{From left to right for the cryptocurrency market index: Hill estimator as a function of $k$; Weissman quantile estimator with D, D-ADJ, and IID 95\% confidence intervals as a function of $k$; LAWS-expectile estimator with D, D-ADJ, and IID 95\% confidence intervals as a function of $k$; QB-expectile estimator with D, D-ADJ, and IID 95\% confidence intervals as a function of $k$.}}
    \label{fig:static_index}
\end{figure}
\begin{table}[!tb]
\centering
\caption{\textit{Composition of the proposed market index of cryptocurrency.}}
\hspace*{-0.6cm}
\begin{tabular}[!ht]{m{6.2em} m{6.2em} m{6.2em} m{6.2em} m{6.2em}}
\hline 
Bitcoin & Ethereum & Litecoin & Monero & Ripple \\
\hline
50\%  & 20\% & 10\% & 10\% & 10\%   \\
\hline
\label{tab:index}
\end{tabular}
\end{table}
 \par
We considered the cryptocurrency market's capitalization and build from scratch a market index with coin weights proportional to the market capitalization of each cryptocurrency. This solution appears to be highly accurate because of the coherence with the definition given by by \cite{acharaya_2017}. The main challenge with this approach was that cryptocurrencies are not stocks, so sometimes the providers of these coins decide to peg the value of these digital currencies to a reference asset. An example is Theter, the third cryptocurrency by capitalization, whose value is fixed at 1 USD. This fact implies that there are just small price fluctuations and thus no tails of the distribution. Looking at the data, we see that Bitcoin has a historical market capitalization consistently above  40\% of the whole crypto market. The second cryptocurrency is Ethereum, with a market capitalization above 18\%. The remaining non-stable coins have a market capitalization under the 2\%, most of them under the 0.05\%. In light of this, it seemed reasonable to us to create an index where Bitcoin and Ethereum had greater importance and divide the rest of the market equally between the remaining cryptocurrencies. Due to the impossibility of obtaining historical data on all cryptocurrencies, the proposed index is composed according to \Cref{tab:index}. We added to our dataset the historical negative log returns of Monero and Ripple, which can be seen as medium-small cap cryptocurrencies such as Litecoin. \par
Recalling the fact that extreme \MES is interpreted as a firm’s propensity
\begin{figure}[!tb]
    \centering
    \hspace*{-.66cm}
    \includegraphics[width=1.08\textwidth,  keepaspectratio]{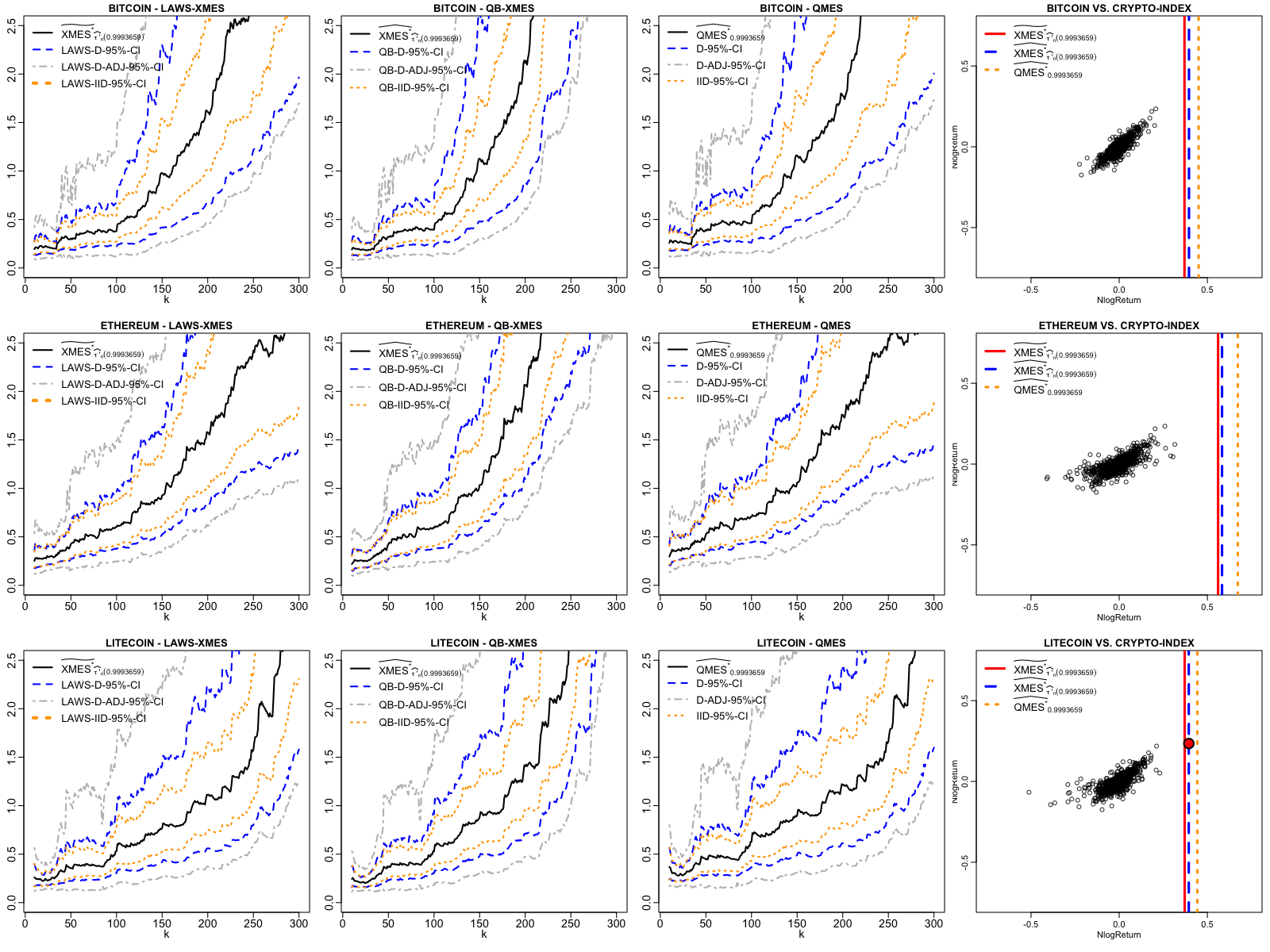}
    \caption{\textit{The first line of the lattice from left to right shows for Bitcoin:  $\XMEStildestar$ with D, D-ADJ, and IID 95\% confidence intervals as a function of k; $\XMEShatstar$ with D, D-ADJ, and IID 95\% confidence intervals as a function of k; $\QMEShatstar$ with D, D-ADJ, and IID 95\% confidence intervals as a function of k; the scatter plot against the cryptocurrency index with the estimates of $\XMEStildestar$,$\XMEShatstar$, and $\QMEShatstar$ with $k=125$. Red dots mark observations that exceed the estimates. The second and third lines show the same charts for Ethereum and Litecoin.}}
    \label{fig:mes}
\end{figure}
to be undercapitalized when the whole market is extremely undercapitalized, it is essential to check that our crypto-market index is effectively heavy-tailed. \Cref{fig:static_index}  summarizes the results. The estimates of the Hill estimators are relatively stable between 80 and 120, with a rough value of 0.4. 
\begin{table}[!ht]
\centering
\caption{\textit{MES estimates with 95\%-D confidence intervals.}}
\begin{tabular}[!ht]{m{8.5em} m{8.5em} m{8.5em} m{8.5em} }
\hline
Estimate & Bitcoin & Ethereum & Litecoin\\
\hline
$\QMEShatstar_{0.9993659}$   & 0.272 [0.450, 0.746]  &  0.427 [0.673, 1.058] & 0.281 [0.443, 0.699]   \\
$\XMEStildestar_{\widehat{\tau'}_n(0.9993659)}$ & 0.223 [0.370, 0.613]  &  0.356 [0.561, 0.882] & 0.236 [0.371, 0.585]   \\
$\XMEShatstar_{\widehat{\tau'}_n(0.9993659)}$   & 0.239 [0.396, 0.656]  &  0.371 [0.583, 0.918] & 0.250 [0.395, 0.622]   \\
\hline
\label{tab:mes}
\end{tabular}
\end{table}
In addition, the charts of the estimates of extreme expectiles and quantiles look fine, with a stable zone between 80 and 130. We can comfortably state that our cryptocurrency market index is heavy-tailed and appropriate for applying our model. \par 
\Cref{fig:mes} shows the result of the analysis. We selected the prudential level $\tau'_n$ using the estimator $\widehat{\tau'}_n(\tau_n)$ with $\tau_n=1-1/n=0.9993659$. The first column of the lattice shows static estimates  $\XMEStildestar$ using the LAWS estimator; the second one shows static estimates  $\XMEShatstar$ using the QB approach; and the last one shows $\QMEShatstar$ estimates calculated using quantiles. All plots show D, D-ADJ and IID confidence intervals. The last column of the lattice of \Cref{fig:mes} exhibits scatter plots of the coins against the cryptocurrency market index with \MES estimates. Red dots in the plots mark observations that exceed the level of estimates. \MES estimates fluctuate as $k$ increases and tend to stabilize in the zone between 80 and 105 for Bitcoin and Ethereum and 100 and 120 for Litecoin. Then, we observe a drift for all the estimates as we start including observations that are not part of the tail of the distribution. Taking $k=100$ can be an acceptable option to make the comparison of the results of the three cryptocurrencies easier. \Cref{tab:mes} summarise the results of the analysis. Interestingly, the confidence intervals of the $\QMEShatstar$ are regularly much wider and uncertain than the ones constructed using expectile-based estimators. \par
Surprisingly, the coin with the highest estimate is Ethereum and not Bitcoin, even if the market index is predominantly composed of the latter. Furthermore, Bitcoin \MES estimates are even lower than Litecoin estimates, even if the last cited contributes to only 10\% of the cryptocurrency market index. There could be many explanations for this phenomenon, which go beyond the scope of this work. One could be that crypto investors in market crashes prefer to shift investments to coins deemed safer, such as Bitcoin. Another answer could be the presence of more loyal investors for Bitcoins, who do not sell in times of crisis in the cryptocurrency market. Nevertheless, we can consider Ethereum as a riskier asset in terms of systemic risk than Bitcoin and Litecoin. 

\section{Conclusions and Further Developments} 
From a theoretical point of view, we have seen that expectiles is the only coherent and elicitable risk measure. In fact, \EVAR is a highly valid alternative to \VAR and \ES, which partially overcame the primary deficiencies of these two traditional risk measures while retaining a valid financial meaning. For this reason, expectile-based risk measures appear to be an attractive tool for future statisticians, asset managers, and in general, for new research in academia. \par
We we have shown the applicability of these tools for the negative log returns of cryptocurrencies. In particular, we have shown that the expectile-based estimators can accurately capture the tail behavior of the returns. Specifically, the \EVAR estimator shows similar behavior to the well-known and celebrated \VAR, which is based on quantile estimates. Moreover, it has been shown that Ethereum exhibits a greater tail risk than Bitcoin and Litecoin, although the marginal tail risk of all three cryptocurrencies should be considered pretty high. \par
From a systemic risk perspective, the \XMES has proven to be a measure capable of evaluating the marginal impact of individual currencies on the crypto market effectively. Again, Ethereum is also systemically riskier than Bitcoin and Litecoin, even if the market capitalization of Bitcoin is more than double that of Ethereum. This counter-intuitive finding is extremely appealing information for both regulators and risk managers, as it implies that in extreme market conditions, some cryptocurrency reacts worse than other. Indeed, \QMES and \XMES can accurately provide point estimates and confidence intervals of this propensity in the context of dependent stationary data, such as the log returns of cryptocurrencies. \par
The results obtained in this work can be further developed in several directions. First, recent academic developments have shown that extreme expectiles can also be efficiently modeled in the context of multivariate time series \citep*{padoan_2022b}. Indeed, enriching the results of this research to time series composed of multiple cryptocurrencies can be very helpful for asset managers operating in the crypto market. Secondly, it could be appealing to perform the same analysis on other cryptocurrencies that have not been taken into account in this work. In addition, the crypto market is a constantly evolving market, where new coins arrive every day, each of them potentially capable of becoming of major importance. \par
Finally, an extension of these results could be assessing systemic risk in the context of a diversified portfolio composed of both cryptocurrencies and other traditional assets, such as stocks and bonds. Cryptocurrencies are already in the portfolios of many investors of various sizes. From an investment perspective, investors are always looking for assets to diversify their risk, following the modern portfolio theory approach. In this portfolio context, cryptocurrencies are an asset class uncorrelated to many others. However, their heavily unstable and emotional nature cannot be fully captured by elementary statistical tools such as the expected returns and the covariance, which are often the only statistical tools used. Because of this, an analysis designed to assess how cryptocurrencies react to severe shocks in the entire financial market could be an incredibly precious asset.

\bibliographystyle{abbrvnat}
\bibliography{bib}
\end{document}